\documentstyle[prd,aps]{revtex}
\begin{document}
\draft

\title{Metrics with distributional curvature}

\author{David Garfinkle
\thanks {Email: garfinkl@oakland.edu}}
\address{
\centerline{Department of Physics, Oakland University,
Rochester, Michigan 48309}}

\maketitle

\null\vspace{-1.75mm}

\begin{abstract}

This paper considers metrics whose curvature tensor makes sense as
a distribution.  A class of such metrics, the regular metrics, was
defined and studied by Geroch and Traschen.  Here, we generalize their
definition to form a wider class: semi-regular metrics.  We then 
examine in detail two metrics that are semi-regular but not regular:
(i) Minkowski spacetime minus a wedge and (ii) a certain traveling 
wave metric.

\end{abstract}


\section{Introduction}

In electrodynamics, one often treats charges and currents that are confined
to points, curves or surfaces.  These idealized sources are supposed to
represent objects where the charge or current is very concentrated.  The
mathematical description for such concentrated sources, as well as the
electric and magnetic fields that they give rise to, is the theory of
distributions.  In electrodynamics, the use of distributions is 
straightforward because Maxwell's equations are linear.  The sum of
distributions or the derivative of a distribution is well defined.
However, in general the product of two distributions is not well defined.
This makes the use of distributions in a nonlinear theory problematic.

In general relativity, one would like to have a treatment of concentrated
sources.  Since the Einstein field equations are nonlinear, one cannot
simply assume that the metric is a distribution, since then the Einstein
tensor would, in general, not be well defined, and the field equations 
would not make sense.  The solution is to identify a class of metrics
whose curvature is well defined as a distribution.  (Alternatively, one
could consider generalized functions\cite{Columbeau1,Columbeau2} instead of
distributions.  This approach has been used;\cite{Vickers1,Vickers2,Vickers3}
but we  will not consider it here).  

For sources on thin shells, an appropriate class of metrics has been
identified.\cite{Israel1,Israel2,Taub2}  Here the class consists of 
metrics that are smooth on
two manifolds with timelike boundary and where the induced metrics on the
two boundaries agree.  Identification of the boundaries then produces a
metric whose curvature makes sense as a distribution.  A similar formalism
can be used to make sense of thin null shells.\cite{Dray,griffiths,b1,b2}

One would like to have a large class of metrics whose curvature makes sense
as a distribution.  This class would include the thin shell metrics, but
would not include all distributional metrics.  Such a class was defined
by Geroch and Traschen\cite{gt} and called regular metrics.  (These 
metrics will be discussed in more detail in the next section).   One of
the results of reference\cite{gt} is that if the curvature of a 
regular metric is concentrated on a surface in spacetime, then the surface
must have dimension 3.  In other words, regular metrics can represent 
thin shells but not thin strings.

Though the class of regular metrics is quite wide, there remains the
possiblility that one can find a wider class.  That is, there may be
metrics that are not regular in the sense of reference\cite{gt} but
whose curvature makes sense as a distribution.  An indication that 
this is the case is provided by an examination of the traveling 
wave metrics.\cite{gv}  These metrics have the form ${{\tilde g}_{ab}} = 
{g_{ab}} + F {k_a} {k_b}$ where $g_{ab}$ is a metric with null,
hypersurface orthogonal Killing vector $k^a$ and $F$ is a scalar.
The expression for the curvature of ${\tilde g}_{ab}$ is linear in $F$.
Thus, provided that $F$ is any scalar that is a distribution, it seems
that the curvature of ${\tilde g}_{ab}$ should  be well defined as a 
distribution.  This is true even for an $F$ for which ${\tilde g}_{ab}$
is not regular.  (One can also make this sort of argument using the 
generalized Kerr-Schild metrics,\cite{Taub} which have the form
 ${{\tilde g}_{ab}} = {g_{ab}} + F {k_a} {k_b}$  
where $k^a$ is a null, shear-free, geodetic congruence on $g_{ab}$).

This paper defines and studies a class of metrics, which we call
semi-regular metrics, which is wider than the class of regular metrics
and whose curvature makes sense as a distribution.  Section 2 gives the
definition of semi-regular metrics.  The next two sections examine
in detail two metrics that are semi-regular but not regular: with
Minkowski spacetime minus a wedge in section 3 and a certain traveling
wave metric in section 4.

\section{Semi-regular metrics}

To define regular metrics, Geroch and Traschen begin with an expression
for the Riemann tensor of a smooth metric.  Let $\nabla _a$ be any
smooth derivative operator with Riemann tensor ${R_{abc}}^d$  and let
${\tilde g}_{ab}$ be any smooth metric.  Define the Christoffel tensor
\begin{equation}
\label{Chr}
{C^c _{ab}} \equiv {1 \over 2} \; {{({{\tilde g}^{-1}})}^{cd}} \, \left ( 
{\nabla _a} {{\tilde g}_{bd}} \, + \, {\nabla _b} {{\tilde g}_{ad}} 
\, - \, {\nabla _d} {{\tilde g}_{ab}} \right ) \; \; \; .
\end{equation}
Then the Riemann tensor of ${\tilde g}_{ab}$ is
\begin{equation}
\label{Riemann}
{{{\tilde R}_{abc}} ^{\;\; \; \; \;  d}} = {{R_{abc}}^d} \; + \; 2 \, {\nabla
_{[b}}  {C^d _{a]c}} \; + \;  2 \, {C^d _{m[b}} {C^m _{a]c}} \; \; \; .
\end{equation}
The idea is to use equation (\ref{Riemann}) as the definition of curvature.
The class of regular metrics is then defined in such a way that the
quantities in equation (\ref{Riemann}) make sense as distributions.  

In this paper we use the definition of tensor distribution given by
Geroch and Traschen. (For a different approach see
reference\cite{tevian2}).  We now recall some of the basic definitions.
(For more detail, see reference\cite{gt}).  A test field is a smooth
tensor density of compact support.  A tensor distribution is a
continuous linear map from the vector space of test fields of a given
index structure to the real numbers.  A tensor field
${\mu _{a \dots c}} ^{b \dots d}$ defined almost everywhere is called
locally integrable if for every test field ${t^{a \dots c}} _{b \dots d}$
the scalar density ${{\mu _{a \dots c}} ^{b \dots d}} {{t^{a \dots c}} _{b
\dots d}}$ is Lebesgue measurable and its Lebesgue integral converges.
Note that through this integral, every locally integrable tensor field
defines a distribution.  A tensor field is called locally square
integrable provided that its outer product with itself is locally
integrable. 
A tensor field ${\mu _{a \dots c}} ^{b \dots d}$ is called locally
bounded provided that the scalar density ${{\mu _{a \dots c}} ^{b \dots
d}} {{t^{a \dots c}} _{b \dots d}} $ is bounded for all test fields
${t^{a \dots c}} _{b \dots d}$.   
For any smooth derivative operator $\nabla _a$
the derivative of a distribution ${S_{a \dots c}} ^{b \dots d}$ is the
distribution ${\nabla _e} {{S_{a \dots c}} ^{b \dots d}}$ such that
\begin{equation}
\label{derivdef}
{\nabla _e} {{S_{a \dots c}} ^{b \dots d}} \left [ {{t^{ea \dots c}} _{b
\dots d}} \right ] 
= - \, {{S_{a \dots c}} ^{b \dots d}} \left [ {\nabla _e} 
{{t^{ea \dots c}}
_{b \dots d}} \right ] 
\end{equation}
where ${t^{ea \dots c}} _{b \dots d}$ is any test field and square
brackets denote the action of the distribution on the test field. 
Let ${{\mu _{a \dots c}} ^{b \dots d}} $ be a locally integrable
tensor field.  Then a locally integrable tensor field ${{W_{ea \dots c}}
^{b \dots d}}$ is called the weak derivative of ${{\mu _{a \dots c}} ^{b
\dots d}}$ provided that 
\begin{equation}
\label{weakdef}
{\int _M} \; {{\mu _{a \dots c}} ^{b \dots d}} {\nabla _e} {{t^{ea \dots
c}} _{b \dots d}} = - \; {\int _M} \; {{W_{ea \dots c}} ^{b \dots d}}
{{t^{ea \dots c}} _{b \dots d}}
\end{equation}
for all test fields ${{t^{ea \dots c}} _{b \dots d}}$.  (Here the
subscript $M$ denotes that the integral is over the whole manifold).
In other words, ${{W_{ea \dots c}} ^{b \dots d}}$ is the locally 
integrable tensor field whose distribution is the derivative of the
distribution ${{\mu _{a \dots c}} ^{b \dots d}}$.  A locally integrable
tensor field may or may not have a weak derivative.

The
definition of regular metric given in reference\cite{gt} is the following:
a metric ${\tilde g}_{ab}$ is called regular provided that (i) 
${\tilde g}_{ab}$ and ${({{\tilde g}^{-1}})}^{ab}$ exist everywhere and
are locally bounded, and (ii) the weak derivative of ${\tilde g}_{ab}$
exists and is locally square integrable.
The idea behind this definition is that we would like $C^c _{ab}$ to be
a tensor field that is both locally integrable and locally square
integrable.  A locally integrable $C^c _{ab}$ defines a distribution,
and since the derivative of a distribution is a distribution, the second
term on the right hand side of equation (\ref{Riemann}) is a distribution.
A locally square integrable $C^c _{ab}$ insures that the third term on
the right hand side of equation (\ref{Riemann}) is locally integrable
and therefore defines a distribution.

These considerations suggest that one could define a larger class of 
metrics with distributional curvature by more directly requiring what
is needed to make the terms on the right hand side of 
equation (\ref{Riemann}) 
distributions.  We therefore define a metric ${\tilde g}_{ab}$ to be
a semi-regular metric provided that (i) ${\tilde g}_{ab}$ and
${({{\tilde g}^{-1}})}^{ab}$ exist almost everywhere and are locally
integrable, and (ii) the weak first derivative of ${\tilde g}_{ab}$
exists and the tensors $C^c _{ab}$ and $ {C^d _{m[b}}{C^m _{a]c}}$
are locally integrable.  

From the definitions, one can show that all regular metrics are
semi-regular, but that some semi-regular metrics are not regular.
In particular, consider the case of a semi-regular metric for which
$C^c _{ab}$ is not locally square integrable.  Then the tensor field
${C^d _{mb}} {C^n _{ac}}$ cannot be regarded as a distribution.
Nonetheless, since ${C^d _{m[b}} {C^m _{a]c}}$ is locally integrable,
there is no obstruction to regarding the right hand side of 
equation (\ref{Riemann}) as a distribution.

Since a locally integrable tensor field defines a distribution, it
follows from the definition that both 
${\tilde R} _{abc} ^{ \; \; \; \; \; d}$
and, by contraction, ${\tilde R}_{ab}$ are distributions for a 
semi-regular metric.  

The class of semi-regular metrics does not include the signature
changing metrics studied by several authors
\cite{ellis1,ellis2,tevian1,hayward1,emb1,tevian3,tevian4,tevian5,tevian6,tevian7,ellis3,ellis4,hayward2,martin,emb2,martin2}
If the signature changing metric is discontinuous, then its weak
derivative does not exist ({\it i.e.} the derivative of the distribution
${\tilde g}_{ab}$ is not a tensor field).  If ${\tilde g}_{ab}$ is 
smooth but degenerate on a hypersurface, then 
${({{\tilde g}^{-1}})}^{ab}$ is not locally integrable. 

For a regular metric, arbitrary products of the 
Riemann tensor, the metric and the inverse metric are distributions.
In particular, this means that the Einstein tensor of a regular metric
is a distribution.  Under what circumstances is this also true of a
semi-regular metric?  For a smooth metric we have
\begin{eqnarray}
\label{Einstein}
{{\tilde G}_{ab}} = {{\tilde R}_{ab}} \; - \; {1 \over 2} \; 
{{({{\tilde g}^{-1}})}^{cd}} {R_{cd}} {{\tilde g}_{ab}} \; + \; 
{{({{\tilde g}^{-1}})}^{cd}} {C^e _{m[c}}{C^m _{e]d}} {{\tilde g}_{ab}}
\nonumber \\
+ \; {\nabla _{[c}} \left ( {C^e _{e]d}} {{({{\tilde g}^{-1}})}^{cd}}
{{\tilde g}_{ab}}\right ) \; + \; {C^e _{d[c}} {\nabla _{e]}} \left (
{{({{\tilde g}^{-1}})}^{cd}} {{\tilde g}_{ab}}\right ) \; \; \; .
\end{eqnarray}

We use equation (\ref{Einstein}) as the definition of the Einstein tensor
for a semi-regular metric.  More precisely, given  a semi-regular
metric ${\tilde g}_{ab}$ we ask whether each term on the right hand side
of equation (\ref{Einstein}) defines a distribution.  If the answer is
yes, then we say that the Einstein tensor of ${\tilde g}_{ab}$ is
given by the right hand side of equation (\ref{Einstein}).  If the answer
is no, then we say that ${\tilde g}_{ab}$ has a Ricci tensor but not an
Einstein tensor.  

In the rest of the paper we will examine two metrics that are semi-regular,
but not regular: Minkowski spacetime minus a wedge and a certain traveling
wave metric.

\section{Minkowski spacetime minus a wedge}

In this section, we treat the metric whose line element is
\begin{equation}
\label{string}
d {{\tilde s}^2} = - \, d {t^2} \; + \; d {z^2} \; + \; d {r^2} \; + 
\; {r^2} \, {\cos ^2} \gamma \, d {\phi ^2} \; \; \; .
\end{equation}
This is the metric of Minkowski spacetime where a wedge of angular
size $\Delta \phi = 2 \pi ( 1 - \cos \gamma )$ has been cut out and
the spacetime has been identified along the cut.  (For a treatment of
this metric in the context of 2+1 dimensional gravity see 
\cite{jackiw1,jackiw2} ).
The metric outside of
a long straight cosmic string is Minkowski spacetime minus a wedge.  
Therefore, it is tempting to regard equation (\ref{string}) as the metric
of a zero thickness cosmic string.  The distributional stress-energy
of ${\tilde g}_{ab}$ (if it is well defined) would then tell us about
the energy content of the string.  In reference \cite{gt} it was shown
that ${\tilde g}_{ab}$ is not regular.  Here, we will show that 
${\tilde g}_{ab}$ is semi-regular, and will calculate its Ricci tensor.

We introduce the ordinary Minkowski metric $g_{ab}$ given by
\begin{equation}
\label{flat}
d {s^2} = - \, d {t^2} \; + \; d {z^2} \; + \; d {r^2} \; + 
\; {r^2} \,  d {\phi ^2} 
\end{equation}
and choose as our smooth derivative operator $\nabla _a$ the one 
compatible with $g_{ab}$.  We lower and raise all tensor indicies with
$g_{ab}$ and its inverse.  Introduce the axial Killing field
$ {\psi ^a} = {{(\partial / \partial \phi )}^a}$.  Then we have
\begin{equation}
\label{string2}
{{\tilde g}_{ab}} = {g_{ab}} \; - \; {\sin ^2} \gamma \, {r^{-2}} \,
{\psi _a} \, {\psi _b} \; \; \; ,
\end{equation}
\begin{equation}
\label{ginverse}
{{({{\tilde g}^{-1}})}^{ab}} = {g^{ab}} \; + \; {\tan ^2} \gamma \,
{r^{- 2}} \, {\psi ^a} \, {\psi ^b} \; \; \; .
\end{equation}
Thus ${\tilde g}_{ab}$ and its inverse are defined almost everywhere.

To find a distribution, one must find its action on test fields: that is
smooth tensor densities of compact support.  
We will write all such
densities as ordinary tensor fields multiplied by the volume element of
$g_{ab}$, and will speak loosely of the ``test tensor field'' $S^{ab}$
when we mean the tensor density ${S^{ab}} {\epsilon _{cdef}}$.  Here,
it is important to keep in mind that the underlying manifold is
$R^4$ with Cartesian coordinates ($t,x,y,z$) where $x = r \cos \phi $
and $y = r \sin \phi $.  Smoothness of a tensor field then means 
smoothness of the Cartesian components as functions of the Cartesian
coordinates.  In particular, ${\psi _a} = x \, {\partial _a} y \, - \,
y \, {\partial _a} x$ so the components of ${r^{-1}} {\psi _a}$
are defined for $r \ne 0$ and are bounded.

For a test field $S^{ab}$, we have
\begin{eqnarray}
\label{string3}
\nonumber
{{\tilde g}_{ab}}{S^{ab}} = S\; - \; {\sin ^2} \gamma \, {r^{-2}} \,
{S^{ab}} {\psi _a} {\psi _b} \\
= S \; - \; {\sin ^2} \gamma \, \left ( {\cos^2} \phi \, {S^{yy}} \,
+ {\sin ^2} \phi {S^{xx}} \, - \, \sin \phi \, \cos \phi \, [
{S^{xy}} \, + \, {S^{yx}} ] \right ) \; \; \; .
\end{eqnarray}
Therefore ${\tilde g}_{ab}$ is locally integrable and thus ${\tilde g}_{ab}$
defines a distribution.  We denote the action of the distribution
${\tilde g}_{ab}$ on the test tensor field $S^{ab}$ by $ {{\tilde g}_{ab}}
[ {S^{ab}}]$.  A similar calculation shows that $ {{({{\tilde
g}^{-1}})}^{ab}}$ is locally integrable.

We now calculate the weak derivative of ${\tilde g}_{ab}$.  That is,
we seek a tensor field $W_{cab}$ defined almost everywhere, such that
\begin{equation}
\label{weakgderiv}
{{\tilde g}_{ab}} [ {\nabla _c} {S^{cab}} ] = - \; {\int _M} \;
{W_{cab}} \, {S^{cab}}
\end{equation}
for all test tensor fields $S^{cab}$.  (Here the subscript $M$ denotes that
the integral is over the whole manifold).  It turns out that the weak 
derivative of ${\tilde g}_{ab}$ is simply ${\nabla _c} {{\tilde g}_{ab}}$
defined at all points where $r \ne 0$.  To see this, note that we have
\begin{eqnarray}
\label{weakg2}
\nonumber
{{\tilde g}_{ab}} [ {\nabla _c} {S^{cab}} ] = {\int _M} \; {{\tilde g}_{ab}}
\, {\nabla _c} \, {S^{cab}} = {\lim _{\epsilon \to 0}} \;
{\int _{r > \epsilon}} \; {{\tilde g}_{ab}}
\, {\nabla _c} \, {S^{cab}} \\
\nonumber
= {\lim _{\epsilon \to 0}} \; {\int _{r > \epsilon}} \; {\nabla _c}
\left ( {{\tilde g}_{ab}} {S^{cab}} \right ) \; - \; {S^{cab}} {\nabla _c}
{{\tilde g}_{ab}} \\
= {\lim _{\epsilon \to 0}} \; \left [ {\int _{r = \epsilon}} \; {n_c}
\, {{\tilde g}_{ab}} \, {S^{cab}} \; - \; {\int _{r > \epsilon}} \;
{S^{cab}}\, {\nabla _c} \, {{\tilde g}_{ab}} \right ] \; \; \; .
\end{eqnarray}
Here, $n_c$ is the unit outward pointing normal to the $r= \epsilon $
surface.

Taking the limit as $\epsilon \to 0$, we find that the first term goes to
zero and the second term becomes an integral over the whole manifold.
Therefore, we have ${W_{cab}} = {\nabla _c} {{\tilde g}_{ab}}$.  However,
we have
\begin{equation}
\label{derivpsi}
{\nabla _a} {\psi _b} = {r^{ - 1}} \; ( {\psi _b} {\nabla _a} r \, - \,
{\psi _a} {\nabla _b} r ) \; \; \; .
\end{equation}
It then follows that
\begin{equation}
\label{derivg}
{\nabla _c} {{\tilde g}_{ab}} = {\sin ^2} \gamma \, {r^{-3}} \, {\psi _c}
\; \left ( {\psi _a} {\nabla _b} r \, + \, {\psi _b} {\nabla _a} r \right )
\; \; \; .
\end{equation}
This quantity is locally integrable, but not locally square integrable.
(Essentially, the reason for this is that ${r^{-1}} {\psi _a}$ is bounded
and $r^{- 1}$ is locally integrable, but $r^{-2}$ is not).  It is this
lack of local square integrability of ${\nabla _c} {{\tilde g}_{ab}}$
that prevents ${\tilde g}_{ab}$ from being a regular metric.

From equations (\ref{Chr}) and (\ref{derivg}) it follows that the
Christoffel tensor is 
\begin{equation}
\label{Chrstring}
{C^c _{ab}} = {\sin ^2} \gamma \, {r^{- 3}} \, {\nabla ^c} r \, {\psi _a}
\, {\psi _b}
\end{equation}
The Christoffel tensor is locally integrable.  Furthermore, since 
${\psi ^a} {\nabla _a} r = 0$, it follows that $ {C^d _{m[b}}{C^m _{a]c}}
= 0 $ and therefore it follows trivially that $ {C^d _{m[b}}{C^m _{a]c}}$
is locally integrable.  It then follows that ${\tilde g}_{ab}$ is a
semi-regular metric.

We now calculate the Ricci tensor of ${\tilde g}_{ab}$.  Contracting 
equation (\ref{Riemann}) we find that for any semi-regular metric
\begin{equation}
\label{Ricci}
{{\tilde R}_{ab}} = {R_{ab}} \; + \; 2 \, {\nabla _{[c}} \, {C^c _{a]b}}
\; + \; 2 \, {C^c _{m[c}} \, {C^m _{a]b}}
\end{equation}
Specializing to the metric of equation (\ref{string}) and using 
${C^c _{cb}} = 0 $ we then find
\begin{equation}
\label{RicciS1}
{{\tilde R}_{ab}} = {\nabla _c} \, {C^c _{ab}} \; \; \; .
\end{equation}
Recall that this is a distributional equation which should be interpreted
as follows: For any test field $S^{ab}$ we have
\begin{equation}
\label{RicciS2}
{{\tilde R}_{ab}} [ {S^{ab}}] = - \, {\int _M} \; {C^c _{ab}} \, {\nabla _c}
\, {S^{ab}} \; \; \; .
\end{equation}
We then find
\begin{eqnarray}
\label{RicciS3}
\nonumber
{{\tilde R}_{ab}} [ {S^{ab}}] = - \, {\lim _{\epsilon \to 0}} \; 
{\int _{r > \epsilon}} \; {C^c _{ab}} \, {\nabla _c} \, {S^{ab}} \\
\nonumber
= - \, {\lim _{\epsilon \to 0}} \; {\int _{r > \epsilon}} \;
{\nabla _c} ( {C^c _{ab}} \, {S^{ab}}) \; - \; {S^{ab}} \, {\nabla _c}
{C^c _{ab}} \\
= - \, {\lim _{\epsilon \to 0}} \; \left [ {\int _{r = \epsilon}} \;
{n_c} \, {C^c _{ab}} \, {S^{ab}} \; - \; {\int _{r > \epsilon}} \;
{S^{ab}} \, {\nabla _c} {C^c _{ab}} \right ] 
\end{eqnarray}
However, for $r \ne 0 $ it follows from equations (\ref{Chrstring}) and
(\ref{derivpsi}) that ${\nabla _c} {C^c _{ab}} = 0$.  Furthermore, we have
${n_c} = - \, {\nabla _c} r$.  Then using equation (\ref{Chrstring}) we find
\begin{eqnarray}
\label{RicciS4}
\nonumber
{{\tilde R}_{ab}} [ {S^{ab}} ] = {\lim _{\epsilon \to 0}} \; 
{\int _{r = \epsilon}} \; {\sin ^2} \gamma \, {r^{-3}} \, {\psi _a} \,
{\psi _b} \, {S^{ab}} \\
\nonumber
= {\sin ^2} \gamma \; {\lim _{r \to 0}} \; {\int _{- \infty} ^ \infty} 
\; d t \; {\int _{- \infty } ^\infty} \; d z \; {\int _0 ^{2 \pi } } \;
d \phi \; \left ( {\sin ^2 } \phi \, {S^{xx}} \; + \; {\cos ^2} \phi \,
{S^{yy}}
\; - \; \sin \phi \, \cos \phi \, ( {S^{xy}} \, + \, {S^{yx}}) \right ) \\
= \pi \, {\sin ^2} \gamma \; {\int _{- \infty} ^ \infty} 
\; d t \; {\int _{- \infty } ^\infty} \; d z \; ({S^{xx}} \, + \, {S^{yy}})
\end{eqnarray}
where the last line is evaluated at $r=0$.  That is, we have
\begin{equation}
\label{Riccistring}
{{\tilde R}_{ab}} = \pi \, {\sin ^2} \gamma \, \delta (x) \, \delta (y) \,
( {\partial _a} x \, {\partial _b} x \, + \, {\partial _a} y \,
{\partial _b} y ) 
\end{equation}

We now calculate the Einstein tensor of ${\tilde g}_{ab}$.  Specializing
equation (\ref{Einstein}) to the metric ${\tilde g}_{ab}$ and using
equations (\ref{derivpsi}), (\ref{derivg}) and (\ref{Chrstring}) we find
\begin{equation}
\label{Einstring}
{{\tilde G}_{ab}} = {{\tilde R}_{ab}} \; - \; {1 \over 2} \; {\nabla _c}
\, \left ( {\tan^2} \gamma \, {r^{-1}} \, {\nabla ^c} r \, {{\tilde g}_{ab}}
\right ) \; \; \; .
\end{equation}
Note that the quantity in parentheses is locally integrable and thus defines
a distribution.  Equation (\ref{Einstring}) therefore makes sense as a
distributional equation whose meaning is that for any test tensor field
$S^{ab}$ we have
\begin{equation}
\label{Einstring2}
{{\tilde G}_{ab}} [ {S^{ab}} ] = {{\tilde R}_{ab}} [ {S^{ab}}] \; + 
\; {1 \over 2} \; {\int _M} \; {\tan ^2} \gamma \; {r^{-1}} \, {\nabla ^c}
r \, {{\tilde g}_{ab}} \, {\nabla _c} {S^{ab}} \; \; \; .
\end{equation}
A calculation analogous to equation (\ref{RicciS4}) then yields
\begin{equation}
\label{Einstring3}
{{\tilde G}_{ab}} = \pi \, {\tan ^2} \gamma \, \delta (x) \, \delta (y)
\; \left [ {\partial _a} t \, {\partial _b} t \; - \; {\partial _a} z 
\, {\partial _b} z \; - \; {1 \over 2} \; {\sin ^2} \gamma \; (
{\partial _a} x \, {\partial _b} x \; + \; {\partial _a} y \,
{\partial _b} y ) \right ] \; \; \; .
\end{equation}
Thus the Einstein tensor of Minkowski spacetime minus a wedge makes sense
as a distribution.

However, in reference \cite{gt} Geroch and Traschen consider one parameter
families of smooth metrics that tend to Minkowski spacetime minus a
wedge.  They show that the limitting value of the mass per unit length
depends not only on the deficit angle, but also on which one parameter
family is used.  Futamase and Garfinkle \cite{gf} consider one parameter
families of self-gravitating Abelian-Higgs strings.  Here too, the
spacetime tends to Minkowski spacetime minus a wedge, and the relation
between mass per unit length and deficit angle depends on which one
parameter family is used.  Thus there is a physically relevant sense
in which Minkowski spacetime minus a wedge does not have a distributional
stress energy.

How are we to reconcile the results of this section with those of 
references \cite{gt} and \cite{gf} on the behavior of one parameter
families of spacetimes tending to Minkowski spacetime minus a wedge?
The answer has to do with the notion of convergence of a one parameter
family of metrics.  Reference \cite{gt} introduces a convergence condition
on a sequence of regular metrics and shows that if a sequence of 
metrics satisfies this convergence condition, then a limit curvature
tensor exists and is the curvature tensor of the limit metric.  If there
is an analogous convergence condition for semi-regular metrics, then
the one parameter families of cosmic strings considered in 
references \cite{gt} and \cite{gf} do not satisfy it.  Thus, though we 
have found a distributional stress energy tensor for Minkowski
spacetime minus a wedge, this distribution may not be an appropriate
physical description of the energy content of the string.

In this section, we have considered only the metric for a long straight
string.  However, the metric of a thin circular loop of cosmic string,
momentarily at rest, has been found by Frolov, Israel and Unruh 
\cite{Unruh1}. Also, a general description of the behavior of curved thin
strings has been given by Unruh, Hayward, Israel and McManus \cite{Unruh2}.  
It would be interesting to see whether these metrics are semi-regular, and 
if so to calculate their distributional curvature.

\section{Traveling wave}

We now consider traveling waves with distributional curvature.  Let
$g_{ab}$ be a smooth metric with null, hypersurface orthogonal Killing
vector $k^a$.  Then there is a scalar $A$ such that ${\nabla _a} {k_b}
= {k_{[b}} {\nabla _{a]}} A$.  Consider the metric ${{\tilde g}_{ab}}
= {g_{ab}} \, + \, F \, {k_a} \, {k_b}$ where $F$ is a scalar satisfying
${k^a} {\nabla _a} F = 0 $.  The physical interpretation of the metric
${\tilde g}_{ab}$ is a wave traveling on the background metric $g_{ab}$
without changing its amplitude or shape.

As shown by Garfinkle and Vachaspati \cite{gv}, the Ricci tensor of the
traveling wave satisfies
\begin{equation}
\label{gvRicci}
{{({{\tilde g}^{-1}})}^{ac}} {{\tilde R}_{cb}} = {{R^a}_b} \; - \; 
{1 \over 2} \; {k^a} \, {k_b} \, {e^{- A}} \, {\nabla _c} {\nabla ^c}
( {e^A} \, F ) \; \; \; .
\end{equation}
Here indicies are lowered and raised with $g_{ab}$ and its inverse.  Since
this expression is linear in $F$, one might hope that the Ricci tensor makes
sense as a distribution in some cases where $F$ is not smooth.

We choose as the background metric $g_{ab}$ for the traveling waves,
Minkowski spacetime expressed as
\begin{equation}
\label{flatuv}
d {s^2} = 2 \, d u \, d v \; + \; d {r^2} \; + \; {r^2} \, d {\phi ^2}
\; \; \; .
\end{equation}
Traveling wave metrics with Minkowski spacetime as a background are
called plane fronted waves.\cite{exact}
In what follows, $g_{ab}$ and its inverse are used to lower and raise
indicies, the volume form of $g_{ab}$ is used to convert tensor fields
into tensor densities and the derivative operator $\nabla _a$ compatible
with $g_{ab}$ is used to define the Christoffel tensor and the curvature
of equations (\ref{Chr}) and (\ref{Riemann}).  The coordinates
($u,v,r,\phi $) are related to the usual Cartesian coordinates 
($t,x,y,z$) by $u = t - z , \; v = - \, (t+z)/2, \; x = r \, \cos \phi $
and $ y = r \, \sin \phi $.  Our null Killing field is ${k_a} = {\nabla _a}
u$.  This $k^a$ is covariantly constant, so $A=0$.  
We consider the traveling wave metric given by
\begin{equation}
\label{travlnr}
{{\tilde g}_{ab}} = {g_{ab}} \; + \; f(u) \, \ln r \, {k_a} \, {k_b}
\end{equation}
where $f(u)$ is any smooth function.

For any test field $S^{ab}$ we have
\begin{equation}
\label{gdotS}
{{\tilde g}_{ab}} {S^{ab}} = S \; + \; \ln r \, f \, {S^{uu}} \; \; \; .
\end{equation}
It then follows that ${\tilde g}_{ab}$ is locally integrable and defines
a distribution.  Paralleling the treatment of equation (\ref{weakg2}) we 
find that the weak derivative of ${\tilde g}_{ab}$ is given for $r \ne 0$
by
\begin{equation}
\label{weakgtrav}
{\nabla _c} {{\tilde g}_{ab}} = \left ( \ln r \, {f '} \, {k_c} \; + \; 
{f \over r} \; {\nabla _c} r \right ) \, {k_a} \, {k_b} \; \; \; .
\end{equation}
Note that this weak derivative is integrable, but not square integrable.
It then follows that ${\tilde g}_{ab}$ is not regular.  (The lack of
regularity of ${\tilde g}_{ab}$ also follows from the fact that
${\tilde g}_{ab}$ is not locally bounded).  

From equations (\ref{Chr}) and (\ref{weakgtrav}) it follows that the
Christoffel tensor is 
\begin{equation}
\label{Chrtrav}
{C^c _{ab}} = {{{f '} \, \ln r}\over 2} \; {k_a} \, {k_b} \, {k^c}
\; + \; {f \over {2 r}} \; \left ( [ {k_a} {\nabla _b} r \, + \,
{k_b} {\nabla _a} r ] \, {k^c} \; - \; {k_a} \, {k_b} \, {\nabla ^c} r 
\right ) \; \; \; .
\end{equation}
This tensor is locally integrable and defines a distribution.  In addition,
${C^c _{cb}} = 0$ and ${C^d _{m[b}} {C^m _{a]c}} = 0 $.  It then follows 
that ${\tilde g}_{ab}$ is semi-regular.

From equation (\ref{Ricci}) and the properties of $C^c _{ab}$ we have
\begin{equation}
\label{Riccitrav}
{{\tilde R}_{ab}} [ {S^{ab}} ] = - \, {C^c _{ab}} \, [ {\nabla _c} 
\, {S^{ab}} ] 
\end{equation}
for any test field $S^{ab}$.  Then a calculation analogous to that of
equations (\ref{RicciS3}) and (\ref{RicciS4}) leads to 
\begin{equation}
\label{Riccitrav2}
{{\tilde R}_{ab}} = \pi \, f \, \delta (x) \, \delta (y) \, {\partial _a}
u \, {\partial _b} u \; \; \; .
\end{equation}
Thus the Ricci tensor is a distribution concentrated at $r=0$.  Using 
equation (\ref{Chrtrav}) in equation(\ref{Einstein}) one can show that
the Einstein tensor is also defined as a distribution, and, in fact, that
${{\tilde G}_{ab}} = {{\tilde R}_{ab}}$.  Therefore, the stress-energy
of this traveling wave is a distribution.

Though we have found a distributional Einstein tensor, and thus, through
Einstein's equation, a distributional stress-energy tensor for a 
semi-regular traveling wave metric, it is not clear what the physical
meaning of this stress-energy tensor is.  To clarify this issue, it 
would be helpful to find a one parameter family of metrics approaching
the metric of equation (\ref{travlnr}) and see whether their Einstein
tensors approach the distribution of equation (\ref{Riccitrav2}).

In this section, we have treated only one traveling wave metric.
However, traveling waves have been found on anti de Sitter 
space, \cite{Siklos} Melvin's Magnetic Universe \cite{gm} and certain
solutions of low energy string theory. \cite{bstrav}  Some of these
metrics are semi-regular, but not regular.  Study of these metrics
could shed further light on the nature of distributional curvature in 
general relativity.

\section{Acknowledgements}

I would like to thank Jennie Traschen for
helpful discussions. I would also like to thank the Institute for
Theoretical Physics at Santa Barbara (partially
supported by NSF grant PHY94-07194) for hospitality.  This work was
partially supported by NSF grant PHY-9722039 to Oakland University.


\begin{references}

\bibitem{Columbeau1}
J. Columbeau, Journal of Mathematical Analysis and Applications,
{\bf 94}, 96 (1983)

\bibitem{Columbeau2}
J. Columbeau, Bulletin of the American Mathematical Society,
{\bf 23}, 251 (1990)

\bibitem{Vickers1}
C. Clarke, J. Vickers and J. Wilson, Class. Quantum Grav. {\bf 13},
2485 (1996)

\bibitem{Vickers2}
J. Vickers and J. Wilson, Class. Quantum Grav. {\bf 16}, 579 (1999)

\bibitem{Vickers3}
J. Vickers and J. Wilson, gr-qc/9807068

\bibitem{Israel1}
W. Israel, Nuovo Cimento B {\bf 44}, 1 (1966)

\bibitem{Israel2}
W. Israel, Phys. Rev. {\bf D15}, 935 (1977)

\bibitem{Taub2}
A. Taub, J. Math. Phys. {\bf 21}, 1423 (1980)
  
\bibitem{Dray}
C. Clarke and T. Dray, Class. Quantum Grav. {\bf 4}, 265 (1987)

\bibitem{griffiths}
J. Griffiths, {\it Colliding Plane Waves in General Relativity}
(Oxford University Press, Oxford 1991)

\bibitem{b1}
C. Barrabes and W. Israel, Phys. Rev. {\bf D43}, 1129 (1991)

\bibitem{b2}
C. Barrabes and P. Hogan, Phys. Rev. {\bf D58}, 044013 (1998)

\bibitem{gt}
R. Geroch and J. Traschen, Phys. Rev. {\bf D36}, 1017 (1987)

\bibitem{gv}
D. Garfinkle and T. Vachaspati, Phys. Rev. {\bf D42}, 1960 (1990)

\bibitem{Taub}
A. Taub, Ann. Phys. (N.Y.) {\bf 134}, 326 (1981)

\bibitem{tevian2}
T. Dray, Int. J. Mod. Phys. {\bf D6}, 717 (1997)

\bibitem{ellis1}
G. Ellis, A. Sumeruk, D. Coule and C. Hellaby, Class. Quantum Grav.
{\bf 9}, 1535 (1992)

\bibitem{ellis2}
G. Ellis, Gen. Rel. Grav. {\bf 24}, 1047 (1992)

\bibitem{tevian1} 
T. Dray, J. Math. Phys. {\bf 37}, 5627 (1996)

\bibitem{hayward1}
S. Hayward, Class. Quantum Grav. {\bf 9}, 1851 (1992). 
Erratum: Class. Quantum Grav. {\bf 9}, 2453 (1992)

\bibitem{emb1}
F. Embacher, Phys. Rev. {\bf D51}, 6764 (1995)

\bibitem{tevian3}
T. Dray, C. Manogue, and R. Tucker, Phys. Rev. {\bf D48}, 2587 (1993)

\bibitem{tevian4}   
C. Hellaby and T. Dray, Phys. Rev. {\bf D49}, 5096 (1995)

\bibitem{tevian5}
T. Dray and C. Hellaby, Phys. Rev. {\bf D52}, 7333 (1995)

\bibitem{tevian6}
T. Dray and C. Hellaby, J. Math. Phys. {\bf 35}, 5922 (1994)

\bibitem{tevian7}
T. Dray, C. Manogue and R. Tucker, Class. Quantum Grav. {\bf 12}, 2767
(1995)

\bibitem{ellis3}
M. Carfora and G. Ellis, Intl. J. Mod. Phys. {\bf D4}, 175 (1995)

\bibitem{ellis4}
T. Dray, G. Ellis, C. Hellaby and C. Manogue, Gen. Rel. Grav. {\bf 29},
591 (1997)

\bibitem{hayward2}
S. Hayward, Phys. Rev. {\bf D52}, 7311 (1995)

\bibitem{martin}
J. Martin, Phys. Rev. {\bf D52}, 6708 (1995)

\bibitem{emb2}
F. Embacher, Class. Quantum. Grav. {\bf 13}, 921 (1996)

\bibitem{martin2}
M. Kriele and J. Martin, Class. Quantum Grav. {\bf 12}, 503 (1995)

\bibitem{jackiw1}
S. Deser, R. Jackiw and G. 'tHooft, Ann. Phys. {\bf 152}, 220 (1984)

\bibitem{jackiw2}
S. Deser and R. Jackiw, Ann. Phys. {\bf 153}, 405 (1984)

\bibitem{gf}
T. Futamase and D. Garfinkle, Phys. Rev. {\bf D37}, 2086 (1988)

\bibitem{Unruh1}
V. Frolov, W. Israel and W. Unruh, Phys. Rev. {\bf D39}, 1084 (1989)

\bibitem{Unruh2}
W. Unruh, G. Hayward, W. Israel and D. McManus, Phys. Rev. Lett.
{\bf 62}, 2897 (1989)

\bibitem{exact}
D. Kramer, H. Stephani, M. MacCallum and E. Herlt, (1980) 
{\it Exact Solutions of Einstein's Field Equations} (Cambridge University
Press, Cambridge).

\bibitem{Siklos}
S. Siklos, (1985) in {\it Galaxies, Axisymmetric Systems and Relativity},
M. MacCallum, Ed. (Cambridge University Press, Cambridge) 

\bibitem{gm}
D. Garfinkle and M. Melvin, Phys. Rev. {\bf D45}, 1188 (1992) 

\bibitem{bstrav}
D. Garfinkle, Phys. Rev. {\bf 46}, 4286 (1992) 

\end{references}
\end{document}